\documentstyle[osa,manuscript,graphicx]{revtex}

\begin{document}

\title{Characterizing teleportation in optics}
\author{T C Ralph, P K Lam and R E S Polkinghorne}
\address{Department of Physics, Faculty of Science, \\ 
         The Australian National University, \\ 
         ACT 0200, Australia \\ 
         Fax: +61 2 6249 0741, \ Telephone: +61 2 6249 2780 \\ 
         e-mail: Timothy.Ralph@anu.edu.au}
\maketitle

\begin{center}
\scriptsize (3rd February 1999)
\end{center}

\begin{abstract}

We discuss the characterization of continuous variable, optical quantum teleportation 
in terms of the two quadrature signal transfer and conditional 
variances between the input and output states. We derive criteria 
which clearly define the classical limits and highlight interesting 
operating points which are not obvious from a calculation of the fidelity 
of the teleportation alone. 

\end{abstract}

\vspace{10 mm}

\section{Introduction}

The quantum state of a system may be transmitted from one location to another through the
direct transmission of only classical information provided, the sender and receiver share a 
non-locally entangled state of the Einstein, Podolsky, Rosen (EPR) 
type \cite {Ein35}. 
This process is called quantum teleportation and was first described in
the context of spin-$1/2$ particles by Bennett et al.~\cite{Ben93}. The first experimental 
realizations \cite{Bou97,bos98} teleported single photon states, albeit with very low
efficiency. More recently Vaidman \cite{Vai94} and Braunstein and Kimble 
\cite{Bra98} have proposed teleportation of 
continuous quantum variables, such as the quadrature amplitudes of the 
electromagnetic field. A high efficiency experimental demonstration of this technique has
been made by Furusawa et al \cite{fur98} using parametric down conversion 
as an EPR source \cite{Ou92}.

Teleportation has applications in quantum computing \cite{bras98} and 
general quantum information manipulation \cite{ben95}. Continuous 
variable teleportation can be applied both through continuous variable 
generalizations of discrete manipulations \cite{bra982} and 
continuous variable manipulations of discrete properties 
\cite{pol98}. Teleportation of optical fields holds great promise due 
to the power of the required optical tools and the maturity of 
relevant optical communications technology.

With these experiments performed and more no doubt planned the issue of how to assess the 
experimental results arises. Traditionally teleportation is quantified by the {\it fidelity}
\cite{sch95} of the process. If the input is in the state $|\phi 
\rangle$ and the output is described
by the density operator $\rho$ then the fidelity is given by 
$F=\langle \phi|\rho|\phi \rangle$. Fidelity is a 
measure of the overlap of the input and output states. In a general experiment one could completely
characterize the input and output states individually via optical 
homodyne tomographic techniques 
\cite{bre97} and thus calculate
the fidelity. However most experimentally realizable optical states 
have Gaussian statistics and can
therefore be fully characterized by measurements of the first and second 
order moments of 
orthogonal quadrature amplitudes,
such as the amplitude and phase quadratures. For a particular set of input states a maximum 
fidelity which can be achieved without the use of entanglement can be derived. For coherent input
states this maximum fidelity is $F=0.5$. This was exceeded in the Furusawa 
et al experiment \cite{fur98}.

On the other hand in the experiments of Boumeester \cite{Bou97} and 
Boschi \cite{bos98} the 
fidelity was very low due to the low efficiency of the teleportation. 
In spite of this (or perhaps, as we shall see,
because of this) strong quantum mechanical correlations in the form of non-local entanglement
were preserved in the teleportation process. This suggests that fidelity alone does not
give the entire picture with regard to the teleportation of particular quantum mechanical
properties. Further, fidelity does not neccessarily recognize the similarity of states which differ only
by reversible transformations. This is important because the desired quantum information
may be preserved in such transformations. If a reversibly transformed version of the input state
emerged from a teleportation process a measure of the fidelity may indicate lack of success
when, in fact, the desired result has been achieved. These examples suggest that additional
measures are required to fully characterize the teleportation process. In this paper we examine
the teleportation criteria proposed by Ralph and Lam \cite{ral98}. 
These criteria are based specifically on the similarity of measurement
results obtained from the input and output of the teleporter, rather than the inferred similarity
of the input and output states. We discuss, with examples, how the above problems are handled by
these criteria.

\section{The Classical Limits}

The criteria are two-dimensional 
and, in analogy 
with the quantum non-demolition measurement (QND) criteria 
\cite{Poi94}, are based on the information transfer 
and quantum correlation achieved by the teleportation. Unlike QND, both 
quadratures of the teleported beam are considered. By examining the limits imposed 
on these quantities in any classical transmission scheme 
criteria for defining quantum teleportation are derived. The criteria are based on 
second order moments of the field which are straightforward to measure. 

In Fig.~1 we show a ``classical teleportation'' scheme. By classical teleportation
we mean that only classical channels are used and no entanglement is involved. An input field, in a 
minimum uncertainty state, is 
detected and the classical information collected is sent to a remote station.  
There the information is used to try to reconstruct the 
original beam. Two possible reconstruction schemes are shown. An 
idealized general method (Fig.1(a)) is to use a
Mach-Zender arrangement with phase modulators in each arm introducing the same transmitted signal
but with a $\pi$ phase shift between them. On the other hand if the input beam is bright, simpler
direct amplitude and phase modulation of a receiver beam can be used 
(Fig.1(b)). In this case the average 
coherent amplitude of the input beam can be regarded as a classical quantity.
We will concentrate on this latter case in developing the criteria. 
We consider an input beam of the form 
\begin{equation}
\hat A_{\rm in}(t) = A_{\rm in}+\delta \hat A_{\rm in}(t)
\label{A}
\end{equation}
where $\hat A_{\rm in}$ is the field annihilation operator; $A_{\rm in}$ is the 
classical, steady state, coherent amplitude of the field (taken to be 
real); and $\delta \hat A_{\rm in}$ is 
a zero-mean operator which carries all the classical and quantum 
fluctuations. For bright beams (i.e. where the classical coherent amplitude is 
much larger than the fluctuations) the amplitude noise spectrum is given by
\begin{equation}
V_{\rm in}^{+}(\omega) = 
\langle |\delta \tilde A_{\rm in}(\omega) + \delta \tilde A_{\rm in}^{\dagger}(\omega)|^{2} \rangle =
\langle |\delta \tilde X_{\rm in}^{+}(\omega)|^{2} \rangle
\end{equation}
where the tilde indicate Fourier transforms have been taken. 
Similarly the phase noise spectrum is given by
\begin{equation}
V_{\rm in}^{-}(\omega) = 
\langle |\delta \tilde A_{\rm in}(\omega) - \delta \tilde A_{\rm in}^{\dagger}(\omega)|^{2} \rangle =
\langle |\delta \tilde X_{\rm in}^{-}(\omega)|^{2} \rangle
\end{equation}
We can write the input light amplitude noise spectrum as $ V_{\rm 
in}^{+} = V_{s}^{+} + V_{n}^{+}$ where 
$V^{+}_{s}$ is the signal power and $V_{n}^{+}$ is 
the noise power.  Similarly the phase noise spectrum can be 
written $V_{\rm in}^{-}=V_{s}^{-}+V_{n}^{-}$. Each Fourier component of the field obeys
the free field boson commutator relation and can thus be considered to be characterizing
 an independent quantum
state. From this point of view we can consider the signal power at a particular 
detection frequency, $\omega$,
as the coherent intensity of
this quantum state and the noise power its quantum fluctuations. In this picture the bright
coherent amplitude of the whole beam serves only as an optical frequency 
and phase reference. It is important to note that this is a formal, 
{\it not physical}, equivalence achieved by going to a rotating 
frame. If we note that $\delta \tilde A^{\dagger}(\omega)=\delta \tilde 
A(-\omega)^{\dagger}$ we see that the spectra are made up of positive 
and negative frequency components. But $\omega$ is the RF frequency 
with respect to the optical carrier, $\omega_{0}$. Hence these positive 
and negative frequency components are really just side-bands of the 
carrier, both at physical (positive) optical frequencies. The properties 
of the observed spectra rely on the characteristics of the photons 
present at both these frequencies. Indeed if one were to put the light 
through a narrow band optical filter centred at $\omega+\omega_{0}$ 
the output light would in general exhibit very different spectra. 
   
Suppose the input light is split into two parts with a 
beamsplitter (see Fig.1(b)).  The amplitude spectrum is detected in one 
arm and the phase spectrum is detected in the other using homodyne 
detection techniques \cite{yue80}. For the case of ideal detection
the following spectra are obtained
\begin{eqnarray}
V_{1}^{+} & = & \eta V_{\rm in}^{+}+(1-\eta)V_{v}^{+} \nonumber\\ 
V_{2}^{-} & = & (1-\eta)V_{\rm in}^{-} + \eta V_{v}^{-}
\label{bs}
\end{eqnarray}
where $\eta$ is the splitting ratio at the beamsplitter.
As the amplitude and phase 
quadratures are conjugate observables it is not possible to obtain 
perfect knowledge of both simultaneously \cite{yam86}. This is ensured by the 
noise penalties, $V_{v}^{+}$ and $V_{v}^{-}$ introduced by the 
beamsplitter.  For the case of only vacuum entering at the empty port 
of the beamsplitter we have $V_{v}^{+}=V_{v}^{-}=1$. By varying the 
beamsplitter ratio we can make either an ideal measurement of the 
phase quadrature or the amplitude quadrature but any simultaneous 
measurement is neccessarily non-ideal. To within a scaling factor 
Eq.~\ref{bs} is a general result which is 
independent of the particular measurement arrangement. It is an 
example of the 
``measurment uncertainty principle'' \cite{art88}.
The measurement uncertainty 
principle is in addition to the intrinsic uncertainty principle 
which requires that $V_{n}^{+}V_{n}^{-}\ge 1$. 
To quantify the measurement limits thus imposed on signals 
we consider the signal transfer coefficients 
of the two quadratures defined by $T^{+} = {\rm SNR}_{1}^{+}/{\rm 
SNR}_{\rm in}^{+}$ for the amplitude quadrature and $T^{-} = {\rm 
SNR}_{2}^{-}/{\rm SNR}_{\rm in}^{-}$ for the phase quadrature.  Here 
${\rm SNR}=V_{s}/V_{n}$ is the signal to noise ratios of the input 
quadratures, ${\rm in}$, and the detected fields, $1,2$.  We find 
quite generally a total transfer coefficient 
\begin{eqnarray}
 T_{q} & = & T^{+}+T^{-}\nonumber\\
& = &{{\eta V_{n}^{+}}\over{\eta V_{n}^{+}+(1-\eta)V_{v}^{+}}}+
 {{(1-\eta)V_{n}^{-}}\over 
 {(1-\eta)V_{n}^{-}+\eta V_{v}^{-}}}\nonumber\\
 & = & 1+{{\eta(1-\eta)(V_{n}^{+}V_{n}^{-}-V_{v}^{+}V_{v}^{-})}\over
 {\eta(1-\eta)V_{n}^{+}V_{n}^{-}+
 \eta^{2} V_{n}^{+}V_{v}^{-}+(1-\eta)^{2}V_{v}^{+}V_{n}^{-}+
 \eta(1-\eta)V_{v}^{+}V_{v}^{-}}}
\end{eqnarray}
We wish to derive a quantum limit so we assume our input beam is in a 
minimum uncertainty state ($V_{n}^{+}V_{n}^{-}=1$). 
Also using the uncertainty relation 
($V_{v}^{+}V_{v}^{-}\ge 1$) we find 
\begin{equation}
 T_{q} \le 1
 \label{info}
\end{equation}
for any simultaneous measurement of both quadratures.  This places an 
absolute upper limit on the signal information that can possibly be 
transmitted through the classical channel.
 
The information arriving at the receiver is imposed on an independent 
beam of light. We now wish to 
consider how well this can be achieved.  The problem is that the light 
beam at the receiver must carry its own quantum noise.  For small 
signals the action of the modulators can be considered additive and we 
will assume that they are ideal in the sense that loss is negligible 
and the phase modulator produces pure phase modulation and similarly 
for the amplitude modulator.  The output field is given by 
 \begin{equation}
 \hat A_{\rm out}=\hat A_a+\delta \hat R_+ + i \delta \hat R_-
 \end{equation}
The fluctuations imposed by the modulators can be written as the 
following convolutions over time \cite{Wis95} 
\begin{eqnarray}
 \delta \hat R_{+} & = &
 \int_{0 }^t {k_{+}(\tau)} 
 {1 \over 2} A_{\rm in} \left( \sqrt{\eta}\delta \hat X_{\rm in}^{+}(t-\tau)+ 
 \sqrt{1-\eta}\delta \hat X_v^{+}(t-\tau) \right) d\tau \nonumber \\
 \delta \hat R_{-} & = & 
 \int_{0 }^t {k_{-}(\tau)} 
 {1 \over 2} A_{\rm in} \left( \sqrt{1-\eta}\delta \hat X_{\rm in}^{-}(t-\tau)+ 
 \sqrt{\eta}\delta \hat X_v^{-}(t-\tau) \right) d\tau
\end{eqnarray}
where $k_{+}$ and $k_{-}$ describe the action of the electronics in 
the amplitude and phase channels respectively.  The amplitude and 
phase quadrature fluctuations of the receiver beam are represented by 
$\delta \hat X_{a}^{+}$ and $\delta \hat X_{a}^{-}$ respectively.  The 
quadrature noise spectra of the output field are 
\begin{equation}
 V_{\rm out}^{+}=V_{a}^{+}+|\lambda_{+}(\omega)|^{2}(\eta V_{\rm 
 in}^{+}+(1-\eta)V_{v}^{+})
\end{equation}
and
\begin{equation}
 V_{\rm out}^{-}=V_{a}^{-}+|\lambda_{-}(\omega)|^{2}((1-\eta)V_{\rm 
 in}^{-}+\eta V_{v}^{-})
\end{equation}
where various parameters have been rolled into the electronic gains, 
$\lambda_{\pm}$, which are proportional to the Fourier transforms of 
$k_{\pm}$.  By making both $|\lambda_{\pm}|^{2}>>1$ the 
signal transfer coefficients for the output, 
$T_{s}^{\pm}={\rm SNR}_{\rm out}^{\pm}/{\rm SNR}_{\rm in}^{\pm}$, can satisfy the 
equality in Eq.~\ref{info}, thus realizing the maximum allowable 
information transfer.  However then the output beam would be much 
noisier than the input beam and hence a very dissimilar state. 
The similarity of the input and output beams can be quantified by 
the amplitude and phase conditional variances \cite{Hol90};
\begin{equation}
 V_{\rm cv}^{\pm} = 
 V_{\rm out}^{\pm}-{{|{\langle {\delta X_{\rm in}^{\pm} \delta 
 X_{\rm out}^{\pm}} \rangle} |^2} \over {V_{\rm in}^{\pm}}}
\end{equation}
The conditional variances measure the amount of independent noise that has been added to the output
quadratures. Amplification or attenuation of noise or signals common with the input do not affect
the conditional variances. This means that reversible transformations, 
i.e. which do not add noise,
such as parametric amplification do not change the conditional variances. 
If $V_{q}=(V_{\rm cv}^{+} + V_{\rm cv}^{-})/2 = 0$ then the 
input and output are maximally 
correlated. Independent coherent beams have $V_{\rm cv}^{+} = V_{\rm cv}^{-} = 
1$, hence $V_{q}=1$.
For our system we find
\begin{equation}
 V_{q} = V_{a}^{+} + V_{a}^{-} + 
 |\lambda_{+}|^{2}V_{v}^{+}+ |\lambda_{-}|^{2} V_{v}^{-}
\end{equation}
Any attempt to suppress the noise penalty in one 
quadrature, say by squeezing the receiver beam, results in a greater 
penalty in the other quadrature. We find 
\begin{equation}
 V_{q}\ge 1
\end{equation}
with the equality obtained for 
$\lambda_{+} = \lambda_{-} = 0$ and a coherent receiver beam. 
That is, the best correlation between the input and output is achieved by not 
transferring any information.  This rather strange result occurs 
because we have already optimized the correlation between input and 
output by choosing a coherent receiver beam.  Any attempt to transfer 
signal information inevitably adds additional uncorrelated noise to 
the output which degrades the correlation. A special case occurs if 
we pick either $\eta=1$ and $\lambda_{-}=0$, or $\eta=0$ and $\lambda_{+}=0$.
That is we choose to only measure and transmit information about one 
quadrature. Then $V_{q}=1$ regardless of the gain used to transmit 
the measured quadrature. We will refer to this as asymmetric 
classical teleportation. 

In principle one could 
measure $V_{\rm cv}^{+}$ directly by performing a perfect QND 
measurement of the amplitude quadrature of the input field and 
electronically subtracting it from an amplitude quadrature measurement 
of the output field.  In a similar way $V_{\rm cv}^{-}$ could in 
principle be measured using a perfect QND measurement of the phase 
quadrature of the input field.  Clearly this is impractical and also assumes that the disturbance
caused to the orthogonal quadrature by the QND measurment of the input
does not change the teleportation process (a valid assumption
for the scheme considered here).  However 
the correlations can be inferred quite easily from individual 
measurements of the transfer coefficients and the absolute noise 
levels of the output field. Suppose the quadrature 
fluctuations of the output field have the form
\begin{equation}
\delta X_{\rm out}^{\pm}=Y^{\pm} \delta X_{\rm in}^{\pm}+ Z^{\pm} \delta 
X_{N}^{\pm}
\end{equation}
where $Y^{\pm}$ and $Z^{\pm}$ are c-numbers and $\delta X_{N}^{\pm}$ 
includes all added noise sources. Then
\begin{equation}
V_{\rm out}^{\pm}=|Y^{\pm}|^{2}V_{\rm in}^{\pm}+|Z^{\pm}|^{2}V_{N}^{\pm}
\end{equation}
and
\begin{equation}
|{\langle {\delta X_{\rm in}^{\pm} \delta 
 X_{\rm out}^{\pm}} \rangle} |^2=|Y^{\pm}|^{2}{V_{\rm in}^{\pm}}^{2}
\end{equation}
so
\begin{equation}
V_{\rm cv}^{\pm}=|Z^{\pm}|^{2}V_{N}^{\pm}
\end{equation}
also
\begin{equation}
T_{s}^{\pm}={{|Y^{\pm}|^{2}}\over{|Y^{\pm}|^{2}V_{in}^{\pm}
+|Z^{\pm}|^{2}V_{N}^{\pm}}}
\end{equation}
and so we find quiet generally that
\begin{equation}
 V_{\rm cv}^{\pm} = (1-T_{s}^{\pm})V_{\rm out}^{\pm} 
\end{equation}

These results are summarized for a coherent input state
in Fig.~2 where $T_{q}$ versus 
$V_{q}$ is plotted as a function of increasing 
gain (we will refer to this as the $T-V$ graph.  The dashed lines represent the limits set by purely 
classical transmission. The dash-double-dot line shows a symmetric 
scheme, i.e. one which detects and transmits information about both 
quadratures equally whilst the solid line is for an 
asymmetric scheme. With symmetric transmission it is only possible to 
reach the classical limits at the extrema of the gain. However, in the 
limit of high gain, asymmetric transmission approaches the point 
$T_{q}=V_{q}=1$. The region between the symmetric, coherent
curve and the classical limits can also be accessed in a 
symmetric transmission scheme with an asymmetric input state such as a squeezed 
state. This is shown as the dot-dashed line in Fig.~2. However for no 
classical detection-transmission scheme or 
input state can one go below 
$V_{q}=1$ or (for a minimum uncertainty state) above 
$T_{q}=1$.    
    
So what do these criteria tell us? The two quadrature transfer coefficient ($T_{q}$) describes the    
reliability with which two independent signal streams which have been encoded simultaneously on the    
conjugate quadratures, can be passed through the classical channel. A quantum channel can carry more     
information than can be reliably extracted from it. In principle all the information carried by a    
classical channel can be extracted, hence information must be lost in going from the quantum to    
the classical channel. If $T_{q}>1$ then the transmission channel can not be considered purely    
classical. The consequences of this with regard to the actual amount of information that can    
be transferred depends on the size and type of the encoded signals. From the point of view that    
the signals are states, $T_{q}$ could be considered a measure of the distinguishability of the    
input states on the output.    
Note again that $T_{q}=1$ is only a quantum limit for minimum uncertainty states.    
    
The limit imposed by the two quadrature conditional variance ($V_{q}$) may be considered     
more fundamental. For two individual beams in any state the limit $V_{q}\ge1$ cannot be exceeded,    
regardless of any common history. By individual we mean that 
independent measurements can be made on each beam. 
If $V_{q}<1$ then it implies that the input and output must be    
considered to be part of the same beam at the quantum level. This represents a neccessary condition    
for the transfer of any quantum correlations between the input and output. Such quantum correlations    
lead to the unique character of quantum information, hence the passing of this limit is very important.    
However most practical applications will also demand good signal to noise transfer.    
    
\section{Quantum Teleportation}         
    
We now introduce a quantum channel and examine under what conditions the classical limits
are exceeded. Consider the electro-optical arrangement that is shown in 
Fig.~3.  It is similar to that employed by Furusawa et al 
Ref.\cite{fur98}.  The entanglement is provided by two coherently related 
amplitude squeezed sources from optical parametric amplifiers 
(OPA's). These are mixed on a 50:50 beam splitter 
(BS1).  The OPA's are seeded with the coherent beams, $\hat v_{1}$ and 
$\hat v_{2}$, giving output beams
\begin{eqnarray}
\hat a & = & \sqrt{H}\hat v_{1}-\sqrt{H-1}\hat v_{1}^{\dagger}\nonumber\\
\hat b & = & \sqrt{H}\hat v_{2}-\sqrt{H-1}\hat v_{2}^{\dagger}
\end{eqnarray}
where $H$ is the parametric gain. They 
are combined with a $\pi/2$ phase shift giving rise to the output beams
\begin{eqnarray}
\hat c & = & \sqrt{H}\hat v_{3}-\sqrt{H-1}\hat v_{4}^{\dagger}\nonumber\\
\hat d & = & \sqrt{H}\hat v_{4}-\sqrt{H-1}\hat v_{3}^{\dagger}
\end{eqnarray}
where $\hat v_{3}={{1}\over{\sqrt{2}}}(\hat v_{1}+i \hat v_{2})$ and 
$\hat v_{4}={{1}\over{\sqrt{2}}}(\hat v_{1}-i \hat v_{2})$, are formally equivalent 
to independent coherent inputs.  
One of the outputs from the beamsplitter ($\hat c$) is sent to 
where we wish to measure the input signal.  There it is mixed with the 
input signal beam on another 50:50 
beamsplitter (BS2). The beams are mixed with 
local oscillators (LO) and homodyne detection is used to measure the 
phase quadrature of one beam and the amplitude quadrature of the 
other (represented schematically in Fig.~3).  The 
photocurrents thus obtained are sent to amplitude and phase modulators 
situated in the other beam ($d$) coming from the mixed squeezed 
sources. Here we attempt to reconstruct the original beam. 
The specific set-up outlined above
could be a convenient experimental realization when using bright beams. 
For dim input beams one could use squeezed vacua to create the 
entanglement and a Mach-Zender modulation scheme (see Fig.1(a)) for 
reconstruction. In general the required
entanglement can be created by the mixing of any two coherently related squeezed beams, irrespective
of the intensity or the quadrature of their squeezing, provided they are mixed with
the correct phase relationship. The following 
results are not dependent on the specific scheme employed.

Following the approach of Ref. \cite{Lam97}, the amplitude and phase noise 
spectra of the output field are found to be
\begin{eqnarray}
 V_{\rm out}^{\pm} & = & {1\over 2} \left| {\sqrt{\eta_{d}} + 
 \lambda_{\pm}\sqrt{\eta_{c} \eta_{e}}} 
 \right|^2 V_a^{\pm} 
+ {1\over 2} \left| {\sqrt{\eta_{d}} - 
 \lambda_{\pm}\sqrt{\eta_{c} \eta_{e}}} 
 \right|^2 V_b^{\mp} \nonumber\\
 & & + \left| {\lambda_{\pm}} 
 \right|^2 \eta_{e} V_{\rm in}^{\pm}+(1-\eta_{d})+\lambda^{2}(1-\eta_{c} 
 \eta_{e})
\label{tele+}
\end{eqnarray}
Here the amplitude (phase) spectra of beams $a$ and $b$ are given by 
$V_{a}^{+}=\sqrt{H}-\sqrt{H-1}$ ($V_{a}^{-}=\sqrt{H}+\sqrt{H-1}$) 
and $V_{b}^{+}=\sqrt{H}-\sqrt{H-1}$ ($V_{b}^{-}=\sqrt{H}+\sqrt{H-1}$) respectively. 
The transmission efficiencies of beams $\hat c$ and $\hat d$ are given by 
$\eta_{c}$ and $\eta_{d}$ 
respectively, whilst the sender's detection efficiency is given by $\eta_{e}$. The 
cross coupling of the phase spectrum of beam $\hat b$ into the 
amplitude spectrum of the output is due to the $\pi/2$ phase shift. We 
assume initially no loss ($\eta_{c}=\eta_{d}=\eta_{e}=1$).  
Consider the situation if beams $\hat b$ and the signal are blocked 
so that just vacuum enters the empty ports of the beamsplitters.  The 
set-up is then just a feedforward loop.  Lam et al \cite{Lam97} have 
shown that the measurement penalty at the feedforward beamsplitter 
(BS1) can be completely canceled by correct choice of the electronic 
gain, allowing noiseless amplification of $V_{a}^{+}$ to be achieved.  
This cancellation can be seen from Eq.~\ref{tele+} with the electronic 
gain set to $\lambda_{+}=1$.  The remaining penalty is due to 
the in-loop beamsplitter (BS2) which, here, is allowing us to detect 
both quadratures.  But now suppose we inject our signal into the empty 
port of the in-loop beamsplitter.  With $\lambda_{+}=1$ we find 
Eq.~\ref{tele+} reduces to 
\begin{equation}
 V_{\rm out}^{+}= 2 V_a^{+} + V_{\rm in}^{+}
\label{tele+1}
\end{equation}
and if beam $a$ is strongly amplitude squeezed such that 
$V_{a}^{+} << 1$ 
then 
\begin{equation}
 V_{\rm out}^{+} \simeq V_{\rm in}^{+}
\label{tele+2}
\end{equation}
Now consider the phase noise spectrum, Eq.~\ref{tele+}. If we impose 
the same electronic gain condition on the fed-forward phase signal as 
we have for the amplitude signal we will get an output spectrum
\begin{equation}
 V_{\rm out}^{-}=2 V_a^{-}+V_{\rm in}^{-}
\label{telew}
\end{equation}
If beam $\hat a$ is strongly amplitude squeezed then the uncertainty 
principle requires $V_{a}^{-} >> 1$ so this is not a useful arrangement.  
However if we perform negative rather than positive feedforward on our 
phase signal such that $\lambda_{-}=-1$ then we will cancel the 
phase noise of beam $\hat a$ and instead see the vacuum noise entering at 
the empty port of the feedforward beamsplitter.  Finally by injecting 
beam $\hat b$ at this port we find 
\begin{equation}
 V_{\rm out}^{-}=2 V_b^{+} + V_{\rm in}^{-}
\label{tele-1}
\end{equation}
Beam $\hat b$ can be made strongly amplitude squeezed without affecting 
Eq.~\ref{tele+2} thus giving us
\begin{equation}
 V_{\rm out}^{-}\simeq V_{\rm in}^{-}
\label{tele-2}
\end{equation}
Hence we have the remarkable result that we can satisfy both 
Eqs.~\ref{tele+2} and \ref{tele-2} simultaneously even though the 
only direct connection between the input and output fields is 
classical; i.e. teleportation of our input field. More generally, 
the spectral variance at some arbitrary quadrature phase angle 
($\theta$) is 
given by
\begin{eqnarray}
 V_{\rm out}^\theta & = & 
 \left\langle {\left| {\delta A_{\rm out}^{\dagger} e^{+i\theta }+ 
 \delta A_{\rm out}e^{-i\theta }} \right|^2} \right\rangle\nonumber\\
 & = & V_{\rm in}^\theta + 2\cos ^2\theta \;V_a^{+} + 2\sin ^2\theta 
 \; V_b^{+}
\end{eqnarray}
This form makes it clear that, provided beam $\hat a$ and beam $\hat b$ are both 
strongly amplitude squeezed, the input and output spectral variances 
will be approximately equal for any arbitrary quadrature angle (not 
just amplitude and phase).  Note that, as 
for other teleportation schemes, no quantum limited information 
about the input field can be obtained from the classical channels. 
This is because it is ``buried'' by 
the large anti-squeezed fluctuations that are mixed with the input 
beam at the measurement site. The strong EPR correlations carried by the 
quantum channel enable this quantum information to be retrieved on the 
receiver beam.

It is clear that under the ideal conditions of no losses and very strong squeezing the
best operating point is unity gain, i.e. $\lambda=1$ where $\lambda_{+}=-\lambda_{-}=\lambda$. 
However this is
not so clear under non-ideal conditions. In 
Fig.~4 we plot $T_{q}$ versus $V_{q}$ for a coherent input as a function of feedforward gain for 
various values of squeezing.  Notice that although a moderate value of 
squeezing allows either information transfer or the correlation to 
be superior to the classical channel limit, squeezing must be greater 
than 50\% before both conditions can be met simultaneously. This limit remains valid for arbitrary 
input states as the point $T_{q}=V_{q}=1$ is the unity gain point for all input states when the
squeezing is 50\%.

We can learn a lot about the operation of the teleporter by looking at the turning points of the 
$T-V$ graph. A maximum in $T_{q}$ occurs for gain 
$\lambda_{G}=(V_{a}+1)/(V_{a}-1)$. 
At this point
the output is simply an amplified version of the input, i.e.
\begin{equation}
\delta \hat A_{{\rm out},G}=\lambda_{G} \delta \hat 
A_{\rm in}+\sqrt{\lambda_{G}^{2}-1}\delta \hat v_{3}^{\dagger}
\end{equation}
where we have subtracted off the classical coherent amplitude of the field as 
per Eq.\ref{A}. On the other hand a minimum in $V_{q}$ occurs for 
gain $\lambda_{\eta}=(V_{a}-1)/(V_{a}+1)=
1/\lambda_{G}$. At this
point the output is simply the attenuated version of the input 
\begin{equation} 
\delta \hat A_{{\rm out},\eta}=\lambda_{\eta} \delta \hat 
A_{\rm in}+\sqrt{1-\lambda_{\eta}^{2}} \delta \hat v_{4}
\end{equation}
The other point of interest is unity gain as this is approximately the point of maximum fidelity
\cite{fur98}. Here the output is formally equivalent to equal amounts of amplification followed by 
attenuation being applied to the input state and is given by
\begin{equation}
\delta \hat A_{{\rm out},F}=\delta \hat 
A_{\rm in}+\sqrt{1-\lambda_{\eta}^{2}}(\delta \hat v_{4}^{\dagger}+\delta 
\hat v_{3})
\end{equation}
In the limit of no squeezing, $V_{a}\to 1$, we go to the classical limit where the 
minimum in $V_{q}$ occurs for $\lambda \to 0$ with the output looking like a 
very strongly attenuated version of the input, and the maximum in 
$T_{q}$ occurs for $\lambda \to \infty$ with the output looking like a 
very strongly amplified version of the input. The output at the unity gain 
point is a very strongly amplifed, then equally strongly attenuated 
version of the input, effectively a classical channel \cite{ral99}. In the opposite limit of 
very strong squeezing, $V_{a} \to 0$, all three points converge on 
unity gain with the effective attenuation and amplification also 
tending to unity, i.e. the output tends to a perfect copy of the input.

As might be expected these ideal results are degraded by loss. The 
effect of loss on the entangled beams is asymmetric, i.e. loss in 
beam $\hat c$ is more detrimental to $V_{q}$ than $T_{q}$, whilst loss in 
beam $\hat d$ is more detrimental to $T_{q}$ than $V_{q}$. The system is 
most vulnerable to detection losses. Detection losses of greater than 
50\% prevent any crossing of the quantum limits. Fortunately, in recent years 
losses in homodyne detection systems have been reduced to under 10\% 
\cite{lam99}.
If beams $\hat a$ and $\hat b$ are not 
minimum uncertainty squeezed states the results will also be degraded 
except at the unity gain point where exact cancellation of the 
conjugate quadrature occurs. 

\section{Discussion}

The best gain to use in a particular teleportation experiment clearly 
depends on the amount of squeezing available and the characteristics 
of the input field you most wish to preserve. For example if the input 
was in a squeezed state and one wished the output to still exhibit 
non-classical statistics you would operate at the point 
$\lambda_{\eta}$. The output will always exhibit some squeezing at 
this point (in the absence of loss), though possibly very small if $\lambda_{\eta}$ is small. 
In fact if the squeezing on the input is very large, squeezing will 
be seen on the output over all gains for which $V_{q}<1$ (irrespective 
of loss), but never 
outside this range. For lower levels of input squeezing the range of 
output squeezing will be reduced. Hence this condition can be 
regarded as neccessary but not sufficient for the teleportation of 
non-classical statistics. We have recently shown that 
very similar behaviour happens in the teleportation of non-local 
correlations \cite{pol98}. One can use a two polarization mode generalization 
of the continuous variable technique to teleport one of a non-locally entangled 
pair of photons. By examining the coincidence detection rates between 
the teleported and unteleported pairs after polarization filtering it 
can be determined if non-local correlations are still present. For the case 
of no loss it is found 
that by sitting at the point $\lambda_{\eta}$ a maximal violation of 
the Clauser-Horne inequality \cite{cla74} is always achieved, 
regardless of the amount of squeezing in the entangled states. This 
occurs because at the point $\lambda_{\eta}$ the output is a pure 
attenuation of the input and nonlocal measures which use coincidence counting 
are insensitive to attenuation. Of course any information carried 
will be strongly degraded if $\lambda_{\eta}$ is too small.

We can now understand the difference in the accomplishments of the photon 
number and continuous variable experiments. In the photon number 
experiment of Boumeester et al \cite{Bou97} the operating 
point was 
effectively a gain of $\lambda_{\eta}$, which was rather small because 
a weakly pumped (thus not very squeezed) type II OPO was used as the source 
of entanglement. This enabled non-local coincidence 
correlations to be shown, 
implying the limit $V_{q}=1$ had been broken, but with very low coincidence count 
rates such that effectively $T_{q}<<1$. Arguing from analogy with the 
results of Ref.~\cite{pol98} the position of this 
experiment on the $T-V$ graph 
might be approximated by the cross on Fig.4. Being a long way 
from unity gain, the average fidelity in this experiment was low. On the other hand in the 
continuous variable experiment of Furukawa et al \cite{fur98} the 
fidelity was maximized by operating at the unity gain point. 
The position of this experiment on the $T-V$ graph is indicated by the star on Fig.4 
which lies under the coherent state limit indicated by the solid line. 
By picking the optimum compromise between signal and correlation 
preservation the greatest similarity between input and 
output states is obtained. However because less than 50\% squeezing 
was available neither 
their signal transfer nor their correlation broke the quantum limits 
individually. We believe that for any useful application of 
teleportation to quantum information manipulation it will be neccessary 
for the teleporter to operate in the lower-right hand 
quadrant of the $T-V$ graph.
 
Discussions with S L Braunstein, H J Kimble and many others helped to clarify 
some of the issues explored in this paper. This work was 
supported by the Australian Research Council.

\begin{figure}
\includegraphics[width=10cm]{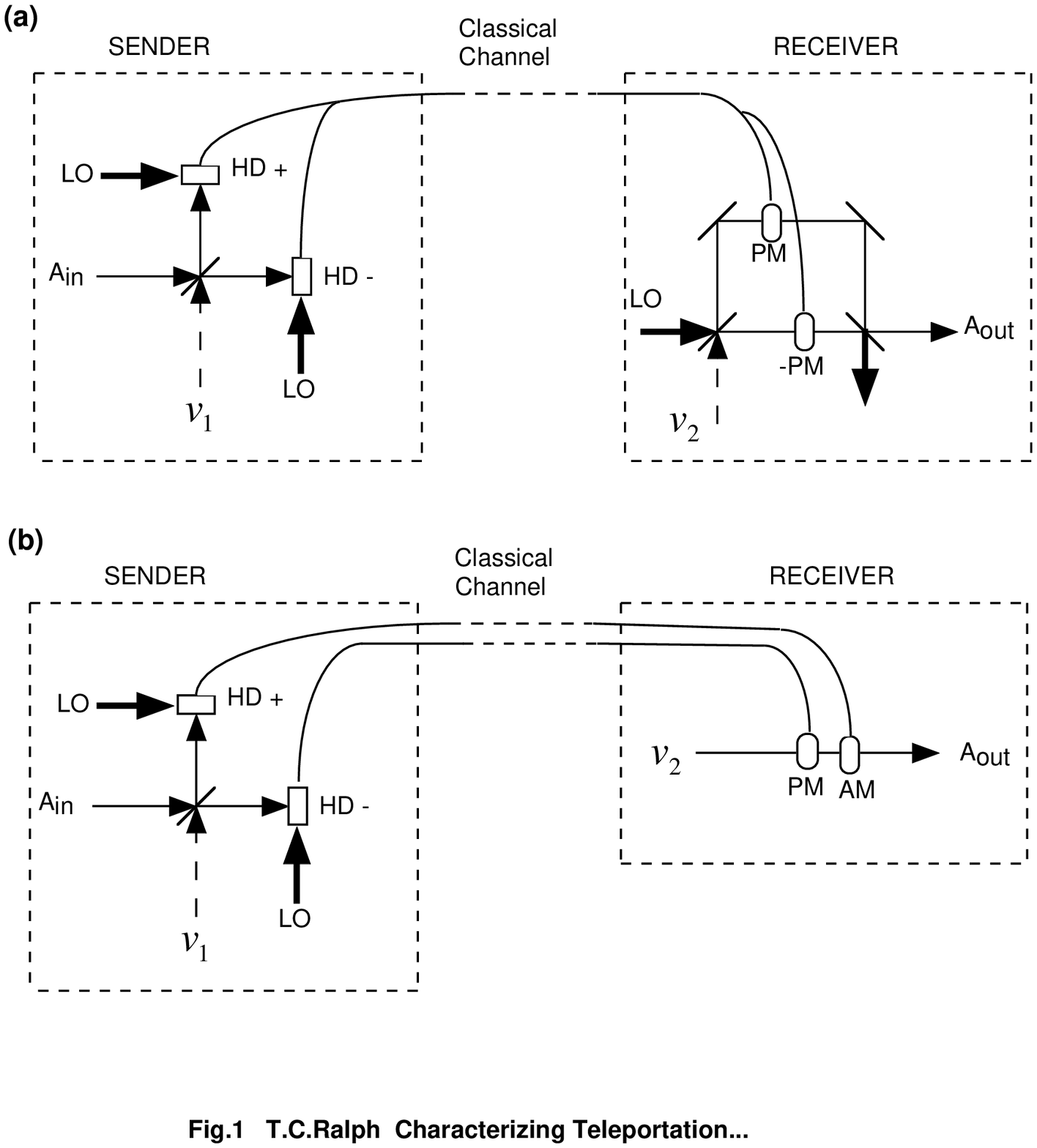}
 \caption{Schematic of classical teleportation arrangements. The 
 arrangement in (a) is an idealized, general one which can in principle cope with 
 any intensity of the input field. The arrangement in (b) can be used 
 if the coherent amplitude of the input beam is large compared to the 
 quantum fluctuations and is the one discussed in detail in this paper. 
 Symbols are; LO:local oscillator, HD+:homodyne 
 detection of the amplitude quadrature, HD-:homodyne detection of the 
 phase quadrature, PM:phase modulation, -PM:phase modulation $\pi$ 
 out of phase, AM:amplitude modulation, $v_{1}$:vacuum input. In (a) 
 $v_{2}$:vacuum input, in (b) $v_{2}$:coherent field with a coherent 
 amplitude equal to that of the input.}
\end{figure}

\begin{figure}
\includegraphics[width=10cm]{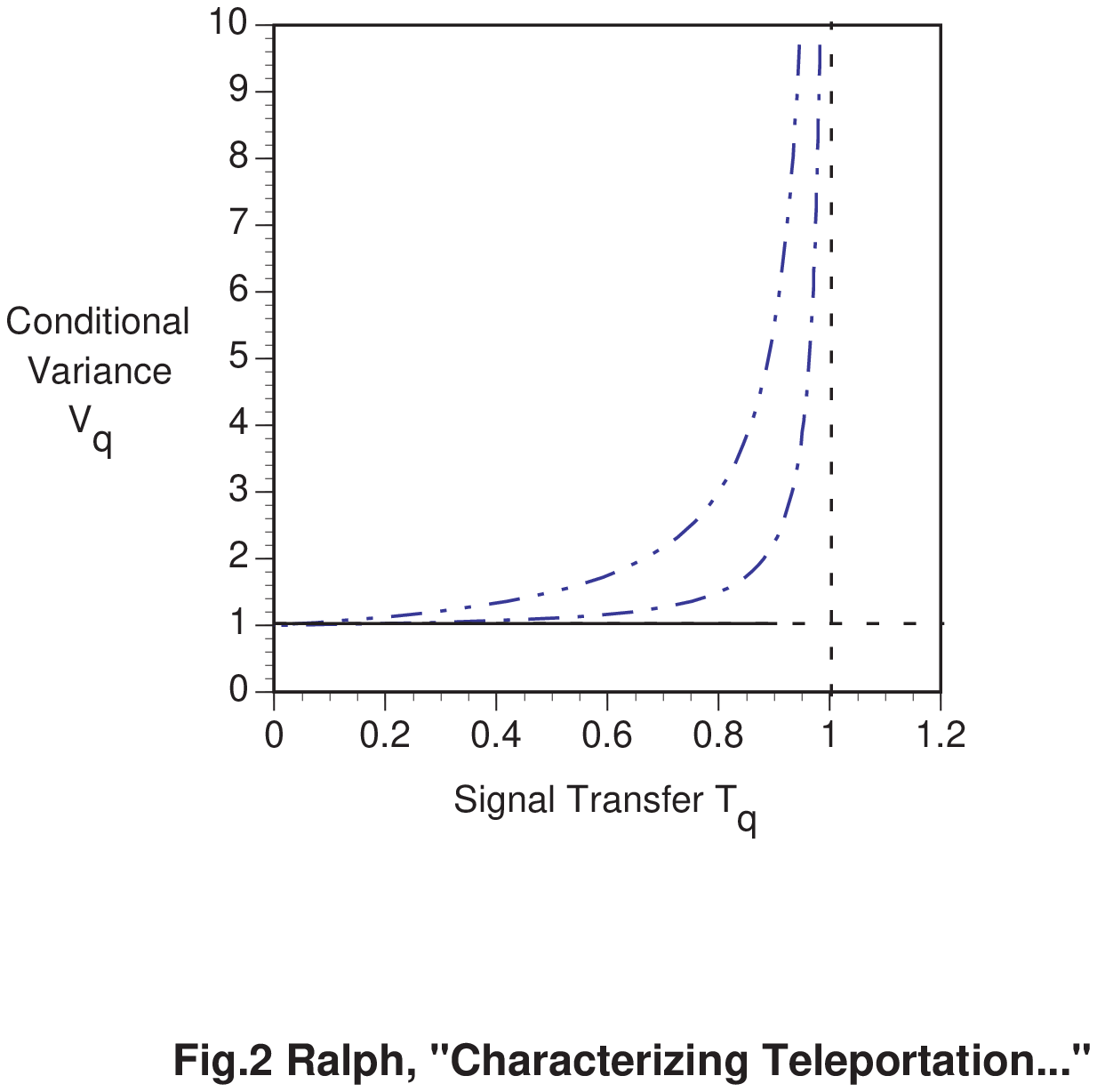}
 \caption{Performance of classical teleportation arrangement.  
 Two quadrature signal transfer ($T_{q}=T_{s}^{+} + T_{s}^{-}$) is plotted versus 
 two quadrataure conditional variance ($V_{q}=(V_{\rm cv}^{+} + V_{\rm 
 cv}^{-})/2$) for 
 $\lambda_{+} = -\lambda_{-}$ running from $0$ to $3.0$.  Dashed lines 
 indicate the classical limits. The double-dot-dashed line is for 
 symmetric transmission of coherent state and the dot-dashed line is 
 for a 90\% squeezed state. The solid line is for asymmetric 
 transmission of a coherent state.}
\end{figure}

\begin{figure}
\includegraphics[width=10cm]{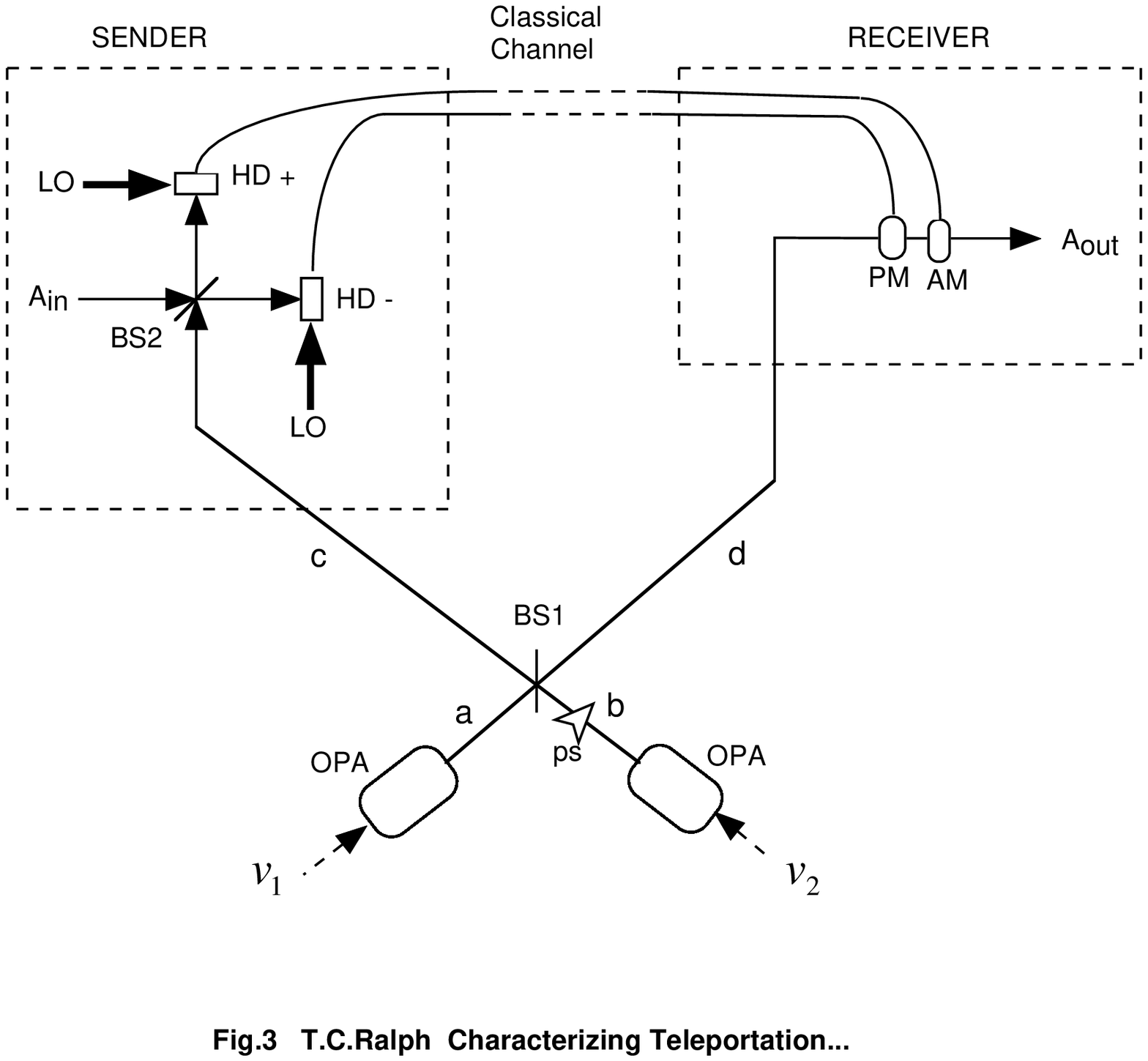}
 \caption{Schematic of quantum teleportation arrangement. Symbols as 
 for Fig.1 plus; OPA:Optical parametric amplifier, ps:$\pi/2$ phase 
 shift and  
 BS1 and BS2 are 50:50 beamsplitters.}
\end{figure}

\begin{figure}
\includegraphics[width=10cm]{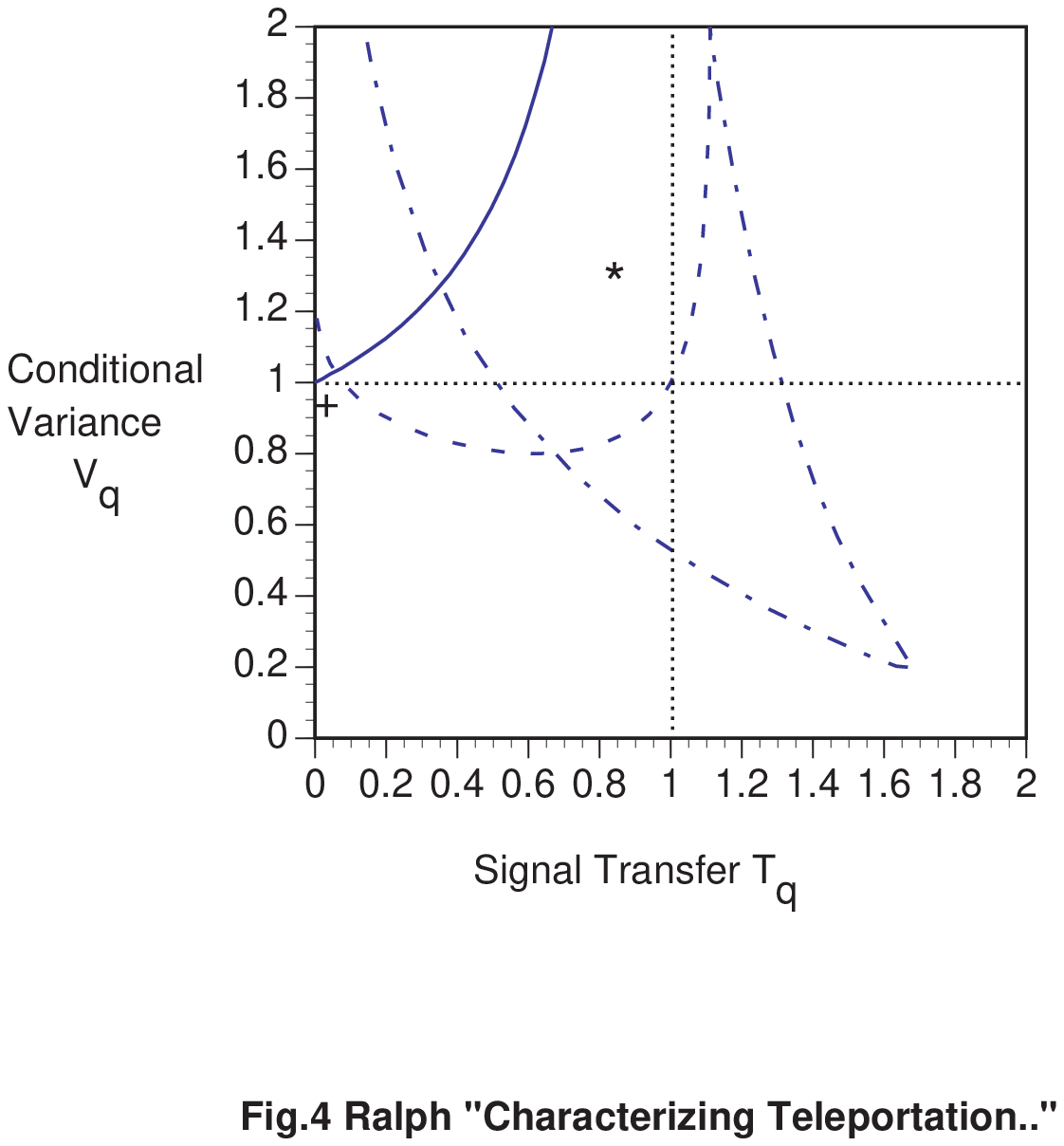}
 \caption{Performance of quantum teleportation arrangement with a 
 coherent input. The solid line is with no squeezing, the dashed line 
 is 50\% squeezing and the dot-dashed line is for 90\% squeezing. The 
 dotted lines are the classical limits for any minimum uncertainty 
 input. }
\end{figure}

\end{document}